\documentclass[
  reprint,
  amsmath,amssymb,
  aps,
  prl,
  superscriptaddress,
  floatfix
]{revtex4-2}

\usepackage{graphicx}      % 插图支持
\usepackage{dcolumn}       % 小数点对齐表格列
\usepackage{bm}            % 数学粗体符号，例如 \bm{k}
\usepackage{mathtools}     % 补充数学功能
\usepackage{braket}        % Dirac 表示
\usepackage{subfigure}     % 并列图
\usepackage{booktabs}      % 高质量表格线
\usepackage{multirow}      % 表格多行合并
\usepackage{xcolor}        % 文字颜色
\usepackage{hyperref}      % 超链接支持
\usepackage{placeins}
\usepackage{float}
\hypersetup{
    colorlinks=true,
    linkcolor=blue,
    citecolor=blue,
    urlcolor=blue
}
\begin{document}

\title{Quantum criticality in a two-component two-dimensional nematic metal}

\author{Yuhang Xiao}
\affiliation{Anhui Province Key Laboratory of Low-Energy Quantum Materials and
Devices, High Magnetic Field Laboratory, HFIPS, Chinese Academy
of Sciences, Hefei, Anhui 230031, China}
\affiliation{Science Island Branch of Graduate School, University of Science and
	Technology of China, Hefei, Anhui 230026, China}
\author{Ning Hao}
\email{haon@hmfl.ac.cn}
\affiliation{Anhui Province Key Laboratory of Low-Energy Quantum Materials and
Devices, High Magnetic Field Laboratory, HFIPS, Chinese Academy
of Sciences, Hefei, Anhui 230031, China}

\begin{abstract}

Nematic fluctuations alter the low-energy electronic behavior and reshape the mean-field phase diagram. We study a two-dimensional nematic metal with a two-component internal degree of freedom. Near the quantum critical point (QCP), longitudinal nematic fluctuations generate the non-Fermi-liquid self-energy $\propto|\omega|^{2/3}$, whereas transverse fluctuations give a subleading logarithmic, marginal-Fermi-liquid correction. Nevertheless, the transverse sector dominates the correction to the phase diagram through a quantum order-by-disorder mechanism: the emergent nematic background reshapes the electronic spectrum, alters the decay of nematic bosons, and produces a negative singular free-energy correction. This correction reconstructs the phase diagram, drives the transition first order, and is described by a scaling function whose two asymptotic limits define two critical regimes with free-energy corrections scaling as $R^{8/3}$ and $R^3$, respectively, where $R$ is the order-parameter magnitude; their powers follow from the competition between the characteristic momentum of normal-state nematic fluctuations and that of the ordered-state background. These two regimes are experimentally distinguishable through nematic-susceptibility hysteresis measurements. The corrected phase diagram shows that the enlarged nodal $\alpha$ phase masks the otherwise bare QCP, making it inaccessible; consequently, the logarithmic self-energy correction is by no means negligible at low energies.

\end{abstract}

\maketitle

% === main text begins ===

\section{Introduction}

Strain experiments on iron pnictides establish nematicity as an electronic instability~\cite{chu2012,kretzschmar2016,kuo2016,dioguardi2016,chen2019,cheng2025}. In two- and three-dimensional Fermi liquids, theory predicts spin-triplet nematic phases with distorted Fermi surface (FS) or winding spin polarization~\cite{wu2007}. The $d$-wave phase with collinear spin polarization provides an interaction-driven route to altermagnetism, combining spin-split bands with zero net magnetization. Photoemission has resolved altermagnetic spin splitting in $\mathrm{MnTe}$~\cite{krempasky2024} and metallic $\mathrm{KV_2Se_2O}$~\cite{jiang2025}.

Near a nematic quantum critical point (QCP), electrons can lose their Fermi-liquid character. Experiments reveal divergent nematic susceptibility without accompanying magnetic order in $\mathrm{FeSe_{1-x}S_x}$~\cite{hosoi2016}, alongside resistivity linear in temperature and anomalous magnetoresistance~\cite{licciardello2019,huang2020,licciardello2019mr,ayres2022}; related transport anomalies occur in $\mathrm{FeSe_xTe_{1-x}}$ films~\cite{sato2025}. For two-dimensional charge nematics, theory obtains a $|\omega|^{2/3}$ electronic self-energy from overdamped fluctuations~\cite{oganesyan2001}. Transverse fluctuations generate logarithmic corrections to the momentum- and frequency-dependent self-energy and interaction vertex~\cite{garst2010}. For spin or pseudospin nematics, electronic coherence must be assessed together with the stability of competing ordered states.

Quantum order-by-disorder (QOBD) describes how quantum fluctuations, through their distinct zero-point energy corrections, select the lowest-energy ordered phase from a ground-state manifold. In three-dimensional spin-triplet nematics, it produces an $M^4\ln|M|$ free-energy correction, drives the uniform transition first order, and favors spiral spin order~\cite{hannappel2016}, where $M$ is a dimensionless order amplitude. Modulated magnetic phases in $\mathrm{PrPtAl}$ support this mechanism~\cite{abduljabbar2015,oneill2021}.

\begin{figure}[t]
  \centering
  \includegraphics[width=0.93\columnwidth]{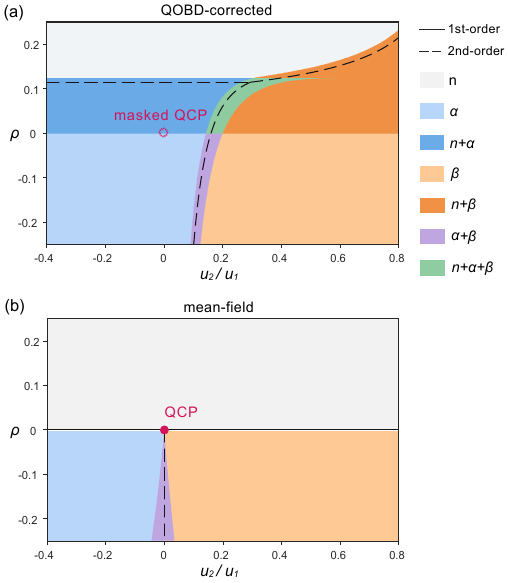}
 \caption{(a) QOBD-corrected phase diagram and (b) uncorrected mean-field phase diagram, both at zero temperature. Here, $n$ denotes the normal state; plus signs indicate that several phases are all local saddle points; and the black dashed line marks the degeneracy of multiple local saddle points, i.e., the first-order transition.}
  \label{Fig1}
\end{figure}

We develop a low-energy theory for quadrupolar ordering in a two-dimensional metal with two spin components or two bands. Coincident and split FS yield distinct nematic dynamics, requiring a nonlocal theory or permitting a local description, respectively. The resulting framework describes both the collinear $\alpha$ phase, with distorted FS, and the $\beta$ phase, with circular contours and winding internal polarization.

With coincident FS, longitudinal nematic fluctuations are overdamped, whereas transverse fluctuations are weakly damped. At one-loop calculation, their contributions to the electronic self-energy scale as $|\omega|^{2/3}$ and $\omega\ln(\omega_0/|\omega|)$, respectively, where $\omega$ is frequency and $\omega_0$ a high-energy scale. The former produces Hertz--Millis-type non-Fermi-liquid behavior~\cite{hertz1976,millis1993,moriya1985,lohneysen2007}; the latter has a marginal-Fermi-liquid form and modifies electronic dispersion and scattering. At the putative QCP, the fractional-power contribution is asymptotically dominant.

Kirkpatrick and Belitz showed that the fluctuation can make spin-nematic transitions first order, explicitly treating a three-dimensional $p$-wave model~\cite{kirkpatrick2011}. Karahasanovic et al. demonstrated that QOBD can generate uniform $d$-wave spin nematicity near a three-dimensional ferromagnetic QCP~\cite{karahasanovic2012}. Starting from a three-dimensional $d$-wave instability, Hannappel et al. considered the $\alpha$ sector, obtaining a quartic logarithmic correction to the free enery and a first-order uniform transition preempted by finite-$\bm{q}$ helical order~\cite{hannappel2016}. We examine uniform $d$-wave order in two dimensions, allowing competition between $\alpha$ and $\beta$. We find that the QOBD correction to the free energy is described by a general scaling function whose two asymptotic limits yield $-R^{8/3}$ and $-R^3$ terms, where $R$ is the order amplitude, defining two critical regimes. These regimes arise from competing fluctuation momenta without and with nematic order and are distinguishable through susceptibility hysteresis. The correction favors $\alpha$ and reconstructs the phase diagram as shown in Fig.~\ref{Fig1}, preempting the bare QCP through a first-order transition. Consequently, the logarithmic electronic self-energy cannot be discarded simply because a stronger singularity exists at the inaccessible QCP.

\section{Effective theory of nematic metal}

We consider a $(2+1)$-dimensional field theory for a two-component Fermi liquid tuned to a Pomeranchuk instability\cite{Pomeranchuk1959,chubukov2018,zacharias2009} in the quadrupolar ($\ell = 2$) channel of an $xy$-(pseudo-)spin interaction. The generalization to higher angular-momentum channels $\ell \neq 2$ is straightforward; we will return to this point below. The low-energy degrees of freedom are fermionic quasiparticles $\psi_\alpha$ and nematic order parameter matrix $\mathcal{Q}_{\mu i}$, which couples to a fermion bilinear $\Phi_{\mu i}$. The Euclidean Lagrangian density is written as
\begin{align}
&\mathcal{L}_0[\psi^\dagger, \psi] = 
\psi^\dagger_\alpha(x)\,\bigl[-G_{\alpha\beta}^{-1}(x)\bigr]\,
\psi_\beta(x), 
\label{L_0 afterHS} \\
&\mathcal{L}_1[\mathcal{Q}] = 
\frac{1}{2}\,\mathcal{Q}_{\mu i}(x)\,\mathcal{Q}_{\mu i}(x), 
\label{L_1 afterHS} \\
&\mathcal{L}_2[\psi^\dagger, \psi, \mathcal{Q}] = 
- \sqrt{g}\,\Phi_{\mu i}(x)\,\mathcal{Q}_{\mu i}(x),
\label{L_2 afterHS}
\end{align}
where repeated indices are summed. Here $\mathcal{L}_0$ describes the underlying Fermi liquid, $\mathcal{L}_1$ encodes the bare stiffness of the nematic field, and $\mathcal{L}_2$ is a Yukawa-type coupling between fermions and nematic fluctuations. The non-interacting fermion propagator is given by
$G^{-1}_{\alpha\beta}(x)= \left[-\partial_\tau - \epsilon(-i\nabla)\right]\delta_{\alpha\beta}
- v_{\mathrm{F}}\Delta\,\sigma^{z}_{\alpha\beta}$,
where $v_{\mathrm{F}}$ is the Fermi velocity and $v_{\mathrm{F}}\Delta$ parameterizes a small band or Zeeman splitting between the two components. The quadratic term $\mathcal{L}_1$ and the fermion–nematic coupling $\mathcal{L}_2$ arise from a Hubbard--Stratonovich decoupling \cite{hubbard1959} of a short-range repulsive interaction of strength $g>0$ in the particle–hole channel.

The Greek indices in the Lagrangian can be interpreted as either real- or pseudo-spin indices. In the real-spin case, $\mathcal{L}_0$ describes a spin-polarized Fermi liquid with a Zeeman splitting $v_{\mathrm{F}} \Delta$. In the pseudo-spin realization, the indices label band, orbital, or other internal components, and $v_{\mathrm{F}} \Delta$ plays the role of a band gap between two otherwise equivalent fermionic species. The fermion bilinear operator in the quadrupolar channel is defined as
\begin{equation}
\Phi_{\mu i}(x) = \psi^\dagger_{\alpha}(x)\, \mathcal{O}^{\mu i}_{\alpha \beta}(-i\nabla)\, \psi_{\beta}(x),
\label{OrderParameter1}
\end{equation}
where $\mathcal{O}^{\mu i}_{\alpha \beta}(-i\nabla) = \sigma^\mu_{\alpha \beta} (-i\partial_a/k_F)\, \tau^i_{ab}\, (-i\partial_b/k_F)$ with $\tau^1=\sigma_z$ and $\tau^2=\sigma_x$.
The Latin index $i$ labels two $d$-wave basis functions, $\hat{\bm{k}}^T \tau^1 \hat{\bm{k}} = \cos(2\theta_{\bm{k}})$ and 
$\hat{\bm{k}}^T \tau^2 \hat{\bm{k}}= \sin(2\theta_{\bm{k}})$, in two dimensions.

The theory defined by Eqs.~\eqref{L_0 afterHS}–\eqref{L_2 afterHS} is invariant under spatial rotations and (pseudo-)spin rotations. For real spin, a finite $v_{\mathrm{F}}\Delta$ explicitly breaks time-reversal symmetry, whereas for pseudo-spin time-reversal symmetry is preserved. The transformation properties of the fermion bilinears under these operations are summarized in the Supplemental Material. In the pseudo-spin realization, an $SO(2)$ rotation corresponds to a relative phase rotation between two bands or orbitals labeled by $\ket{1}$  and  $\ket{2}$. This is captured by the complex order parameter $z_i \equiv \Phi_{x i} + i \Phi_{y i}$, whose expectation value reflects the relative phase structure of multi-band wave functions $\ket{\psi} =  |a| e^{i\phi_1} \ket{1} +  |b| e^{i\phi_2} \ket{2}$, as illustrated in Eq. (\ref{z_i}).

\begin{equation}
\langle z_i \rangle\sim \bra{\psi} \sigma^x \ket{\psi} + i \bra{\psi} \sigma^y \ket{\psi} \sim |a||b|\, e^{i(\phi_2 - \phi_1)}.
\label{z_i}
\end{equation}

Integrating out the fermions generates an effective theory for the nematic order parameter. At this stage it is crucial to distinguish two physically distinct situations, depending on whether the electronic spectrum is already gapped before nematic long-range order sets in. The corresponding effective descriptions are qualitatively different. In the gapless case, where the FS associated with the two components coincide, there is no intrinsic low-energy scale. As a result, integrating out the fermions produces a highly nonlocal kernel, and the effective action cannot be cast into a simple local Landau functional. In this regime the dynamical part of the fermion bubble takes the form
\begin{equation}
\chi_{\mu i, \nu j}(\bm{q}, q_n)
=
-2 \bar{\nu}\, \delta_{\mu \nu}
\int \frac{\mathrm{d}\theta_{\bm{k}}}{2\pi}
\frac{iq_n}{iq_n - v_{\mathrm{F}}\, \hat{\bm{k}} \cdot \bm{q}}\,
e_{\bm{k}}^i e_{\bm{k}}^j,
\label{chi for gapless case}   
\end{equation}
where $q_n\neq 0$, $\bar{\nu}$ is the (dimensionless) angularly averaged density of states near the FS, and $e_{\bm{k}}^{i}= k_a \tau^i_{ab} k_b$. The nonlocal structure encoded in Eq.~\eqref{chi for gapless case} is the origin of the singular self-energies and quantum-critical transport discussed below.

In contrast, when a gap is present between the two components, we may focus on the low-energy regime where $\left|\bm{q}\right| \ll \Delta$ and $q_n / v_{\mathrm{F}} \ll \Delta$. The gapped fermions act as short-ranged mediators, and the effective theory becomes local, as given by
\begin{align}
\mathcal{L}_{\text{eff}} 
&= -\frac{i \alpha}{2}\, \epsilon_{\mu \nu}\, \mathcal{Q}_{\mu i}\, \partial_\tau \mathcal{Q}_{\nu i}
+ \frac{\kappa}{2}\, \nabla \mathcal{Q}_{\mu i} \cdot \nabla \mathcal{Q}_{\mu i}
+ \frac{\rho}{2}\, \mathcal{Q}_{\mu i} \mathcal{Q}_{\mu i} \nonumber \\
&\quad + \frac{u_1}{4} \left( \mathcal{Q}_{\mu i} \mathcal{Q}_{\mu i} \right)^2
- \frac{u_2}{4} \left( \epsilon_{\mu \nu} \epsilon_{i j} \mathcal{Q}_{\mu i} \mathcal{Q}_{\nu j} \right)^2,
\label{L_eff}
\end{align}
where the coefficients are determined by the density of states $N(\epsilon)$ and the microscopic coupling $g$. Explicit expressions for $\alpha$, $\kappa$, $\rho$, $u_1$, and $u_2$ are given in the Supplemental Material.

The first term in Eq.~\eqref{L_eff} reflects a Berry-phase-like dynamics and implies a dynamical exponent $z=2$. An additional subtlety arises in systems with a fixed electron density. In this case the quartic coefficient $u_1$ acquires an extra contribution of the form $g\,[N^{\prime}(0)]^2/N(0)$, which can significantly enhance $u_1$ and thereby stabilize nematic order. This structure is natural from a thermodynamic perspective: keeping the density fixed requires the quadratic coefficient $\rho$ to shift so as to compensate the change in electron density induced by $\langle\mathcal{Q}\rangle\neq 0$. Since the density shift is of order $\mathcal{Q}^2$, the feedback into the free energy appears at order $\mathcal{Q}^4$, and is thus encoded in quartic term.

\begin{figure}[t]
  \centering
  \includegraphics[width=0.95\columnwidth]{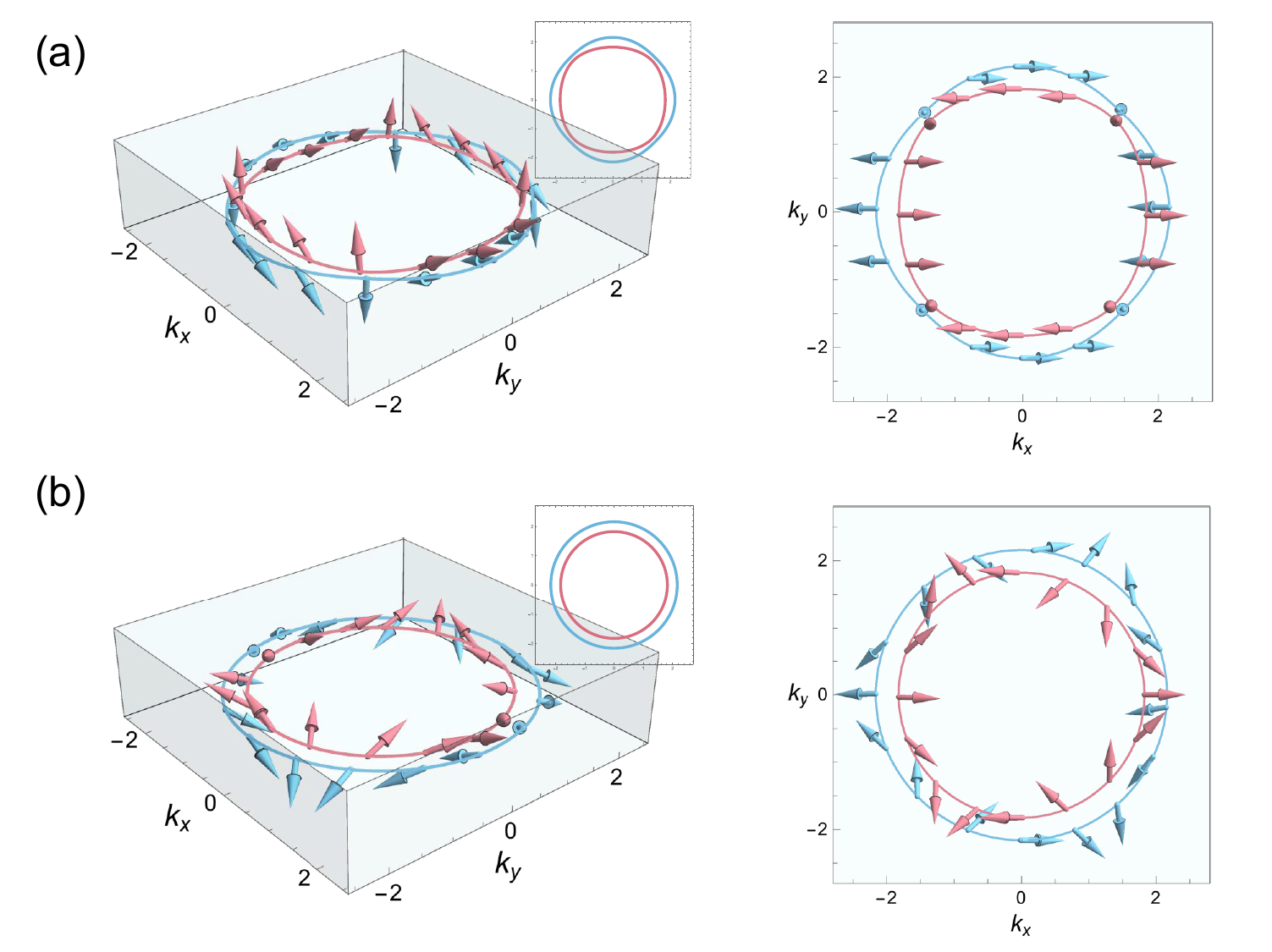}%
  \caption{(a) and (b) correspond to the $\alpha$ phase and $\beta$ phase, respectively, with $\ell = 2$ and $v_{\mathrm{F}} \Delta \neq 0$. The right panels show the in-plane projections of the (pseudo-)spin textures.
}
  \label{fig:spin textures}
\end{figure}

\section{Ordered phases}

After minimizing Eq.~\eqref{L_eff}, we obtain two distinct mean-field solutions, which we refer to as the $\alpha$ and $\beta$ phases. Writing $\mathcal{Q}_i=(\mathcal{Q}_{xi},\mathcal{Q}_{yi})$, the $\alpha$ phase is characterized by $\mathcal{Q}_1 \parallel \mathcal{Q}_2$, whereas in the $\beta$ phase one has $\mathcal{Q}_1 \perp \mathcal{Q}_2$. The corresponding Fermi-surface geometries and internal textures for the $\ell=2$ channel are illustrated in Fig.~\ref{fig:spin textures}. In the $\alpha$ phase, the FS is distorted and the in-plane (pseudo-)spin polarization is collinear. In the $\beta$ phase, by contrast, the FS remains isotropic; the symmetry breaking is encoded entirely in a momentum-space texture of the internal two-component degree of freedom with a $\pm \ell$ winding. In both phases, the net dipole moment of the internal degree of freedom vanishes.  Notably, phases $\alpha$ and $\beta$ remain robust regardless of whether $\Delta=0$ or  $\Delta \neq0$.

Interestingly, for real spin, the $\alpha$ phase in the $\ell=2$ channel at $\Delta=0$ belongs to the broader family of unconventional magnetism~\cite{liu2025}, classified by the angular structure of their spin order in analogy with unconventional superconductivity. In particular, its $d$-wave spin splitting and zero net magnetization closely resemble collinear altermagnetism. Our continuum model generates this texture through a spontaneous quadrupolar instability of an initially spin-degenerate metal. In crystalline altermagnets, lattice and magnetic symmetries further constrain the spin pattern, and an itinerant origin has also been proposed~\cite{qian2025}. Moreover, the nematic spin splitting represents an interaction-generated coupling between spin and momentum, closely related to the broader concept of spontaneous spin--orbit coupling (SOC)~\cite{wu2004,raghu2008,zhang2025phonon}. In the presence of bare SOC, electronic ordering can further reshape the spin texture and support a variety of symmetry-breaking states~\cite{fu2015,yuan2026}.

\section{Critical behavior of bosons}

The NFL properties of the electrons originate from the critical dynamics of the nematic fluctuations, which act as the effective glue mediating singular electron–electron scattering. In the disordered phase, where long-range nematic order is absent, the soft modes are strongly damped by particle–hole excitations. Their general structure has been discussed previously \cite{oganesyan2001,wu2007}; here we briefly summarize the key ingredients and present the form of the bosonic self-energy relevant for our analysis.

At this stage, we focus on the gapless case where the two FS coincide. The dynamical exponent $z=2$ discussed below refers to the underdamped transverse nematic mode emerging from the nonlocal Landau-damped kernel in the rotated momentum-space basis, and should be distinguished from the Berry-phase-like $z=2$ dynamics that arises in the local effective theory when a gap is present between the two components.
Since the absence of a gap allows the fermions on the two components to participate equally in the low-energy processes that determine the nematic self-energy. By introducing a $\bm{q}$-dependent basis $\{\hat{q}',\hat{q}'_t\}$, Eq.~\eqref{chi for gapless case} becomes diagonal and takes the form
\begin{equation}
\hat{\chi}_{\mu\nu}(\bm{q},q_n)
= 2\bar{\nu}\,\delta_{\mu\nu}\times
\left\{
\begin{array}{ll}
|x|\,\hat{q}'\hat{q}'
+ \ell x^2\,\hat{q}'_t\hat{q}'_t, & \text{even }\ell \\[8pt]
\ell x^2\,\hat{q}'\hat{q}'
+ |x|\,\hat{q}'_t\hat{q}'_t, & \text{odd }\ell
\end{array}
\right.,
\label{result of chi}
\end{equation}
where $x=q_n/(v_{\mathrm{F}}|\bm{q}|)$ is the natural scaling variable.

The exponent of $x$ in Eq.~\eqref{result of chi}, denoted by $\gamma$, controls the damping of the nematic modes and fixes the dynamical exponent via $z=1+2/\gamma$ ($\gamma>0$). In $d+1$ dimensions, the real-space propagator of a soft mode assumes the scaling form
\begin{equation}
D(\bm{r},t)\sim \frac{1}{r^{d+z-2}}\,
\Psi\!\left(t/r^z\right),
\label{propagator in real space}
\end{equation}
and the value of $\gamma$ dictates how the fluctuation spreads and decays. The qualitatively distinct propagation behaviors are summarized in Table~\ref{Tab:propagators_for_different_z}.

\begin{table}[t]
\caption{
Different scaling exponents lead to distinct propagation behaviors. $\gamma=0, 1, 2$ correspond to undamped, Landau-damped and underdamped behaviors, respectively.
Here, $\omega_0 = c/r^2$ and $\xi = c^{-1}$.
}
\centering
\renewcommand{\arraystretch}{1.4}
\begin{tabular}{c @{\hspace{20pt}} c @{\hspace{25pt}} c @{\hspace{25pt}} c}
\hline\hline
$\gamma$ & $z$ & $D^{-1}(\Omega, \bm{q})$ & $D(t,\bm{r})$ at large $t$ \\
\hline
0 & 1 & $c^{-2}\Omega^2 +  q^2$ 
  & $\dfrac{1}{r} e^{-r/\xi} \cos(\omega_0 t)$ \\
1 & 3 & $c^{-2}\Omega^2 +  q^2 +  \dfrac{|\Omega|}{q}$ 
  & $\dfrac{1}{r^3} e^{-t/r^3}$ \\
2 & 2 & $c^{-2}\Omega^2 +  q^2 +  \dfrac{\Omega^2}{q^2}$ 
  & $\dfrac{1}{r^2} e^{-t/r^2} \cos(\omega_0 t)$ \\
\hline\hline
\end{tabular}
\label{Tab:propagators_for_different_z}
\end{table}

Eq.~\eqref{propagator in real space} reveals a striking alternation in the damping pattern controlled by the angular-momentum channel. For even $\ell$, the mode polarized along $\hat{q}'$ is overdamped with $z=3$, while its transverse partner is underdamped with $z=2$; for odd $\ell$, the roles of the two modes are interchanged. The distinction here is geometric: “longitudinal’’ and “transverse’’ are defined with respect to the rotated basis $\{\hat{q}'=(\cos(\ell\theta_q),\sin(\ell\theta_q)),\hat{q}'_t=(-\sin(\ell\theta_q),\cos(\ell\theta_q))\}$ rather than the physical momentum direction $\bm{q}$. As the momentum varies, the orientation of $\bm{\hat{q}}'$ winds relative to $\bm{q}$ with a vorticity set by $\ell$, imprinting an angular-momentum–dependent texture onto the critical nematic fluctuations.

The origin of the difference of damping behaviors is directly kinematic. For example, for $\ell=2$, a low-energy particle--hole pair with small momentum $\bm q$ lies near a point where $\bm q$ is tangent to the FS, since its excitation energy is $v_Fq\cos(\theta_k-\theta_q)$. At that point the longitudinal vertex $\cos[2(\theta_k-\theta_q)]$ is nonzero, whereas the transverse vertex $\sin[2(\theta_k-\theta_q)]$ vanishes. The transverse mode therefore has less access to the particle--hole continuum.

\section{Critical behavior of electrons}

Having characterized the damping of the nematic modes arising from their coupling to particle–hole excitations, we now examine the converse influence of these soft fluctuations on the electronic degrees of freedom. The critical modes scatter fermions over the entire FS, and the resulting one-loop self-energy takes the form
\begin{equation} 
\Sigma_{\alpha \beta}(k) 
= \frac{g}{k_F^4} 
\int \frac{\mathrm{d}^3 q}{(2\pi)^3} 
e_{\bm{k}}^i \, e_{\bm{k}}^j \, 
\sigma_{\alpha \gamma}^{\mu}\sigma_{\gamma \beta}^{\nu}\,
D_{\mu i, \nu j}(q)\,
G(k+q),
\label{definition of Sigma }
\end{equation}
where $D_{\mu i,\nu j}$ denotes the propagator of the nematic bosons, given by $D_{ij}^{-1} = \bm{q}^2 + c^{-2} q_n^2 + \chi_{ij}(q)$ with $\chi_{ij}$ encoding the Landau damping effects discussed above. We have included a bare part $D_0^{-1} = \bm{q}^2 + c^{-2} q_n^2$ in the propagator (henceforth, we set $c=1$). Although this bare contribution does not follow from a strictly local HS decoupling, it naturally appears once the microscopic interaction has any momentum or frequency structure.

\begin{figure}[t]
  \centering
  \includegraphics[width=0.98\columnwidth]{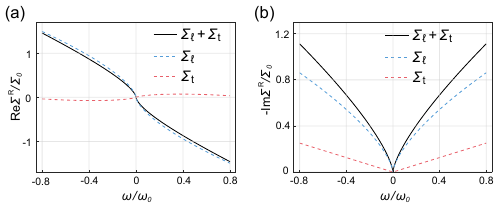}%
  \caption{(a) Real and (b) imaginary parts of the non-Fermi-liquid self-energy. The real part governs energy renormalization, and the imaginary part determines the quasiparticle scattering rate. The self-energy constant $\Sigma_0$ used for normalization in the figure depends on the energy cutoff $\omega_0$ and other microscopic parameters.}
  \label{Fig3}
\end{figure}

Fixing the momentum on the FS and focusing on the frequency dependence, the critical fermionic self-energy separates into two contributions, $\Sigma_{\ell}$ and $\Sigma_t$, arising from scattering off longitudinal and transverse nematic modes, respectively (here we set $\ell = 2$ for concreteness). Evaluating the one-loop diagrams  and omitting overall numerical prefactors, we obtain
\begin{align}
\Sigma_\ell(i\omega)
 &\sim i \,\mathrm{sgn}(\omega)\,|\omega|^{2/3}
       +o(|\omega|^{2/3}),\label{Sigma_l}\\
\Sigma_t(i\omega)
 &\sim i \,\omega\ln\frac{\omega_0}{|\omega|}
       +o(i\omega).
\label{Sigma_t}
\end{align}
where $\omega_0$ is a fixed ultraviolet matching scale. The transverse channel supplies a logarithmic correction to the frequency and dispersion dependence rather than a stronger fractional power. Longitudinal fluctuations thus produce the leading departure from Fermi-liquid behavior. The real and imaginary parts of the non-Fermi-liquid self-energy are shown in Fig.~\ref{Fig3}, with the real part of the logarithmic contribution remaining logarithmic and its imaginary part linear.

At finite temperature or finite frequency, the self-energy arising from logarithmic transverse fluctuations still substantially affects the electronic spectral function. Even at low energies, although the critical self-energy remains of the conventional Hertz--Millis type ~\cite{hertz1976,millis1993,moriya1985}, our calculations in the next section show that, in fact, the system cannot reach the QCP; before doing so, it has already undergone a first-order transition. Consequently, the self-energy contributions from transverse and longitudinal fluctuations are of nearly equal importance at low energies.

\section{Fluctuation-induced ordering}

QOBD describes how the zero-point energy of quantum fluctuations selects among competing mean-field configurations. States that are degenerate at the mean-field level can support different fluctuation spectra and therefore acquire different quantum corrections. These corrections reshape the energy landscape on the mean-field ground-state manifold and alter its competition with nearby configurations, selecting the ground state and potentially reconstructing the mean-field phase diagram. In a nematic metal, this mechanism involves a feedback: each ordered configuration changes the electronic spectrum, which in turn modifies the dynamics and zero-point contribution of the nematic fluctuations. To calculate this correction, we prescribe a uniform nematic background, integrate out the fermions, and expand the resulting effective action to quadratic order in nematic fluctuations. Integrating over these Gaussian fluctuations gives a background-dependent free-energy correction.

The background produces an internal field $\bm d(\theta_{\bm{k}})=\mathcal{Q}\bm {e}_{\theta_{\bm{k}}}$ and two bands $\epsilon_{\bm k}\pm|\bm d(\theta)|$. Here, $\mathcal{Q}$ is a tensor; its product with the real-space vector $\bm{e}_\theta$ yields a vector $\bm{d}$ residing in the internal-degree-of-freedom space. A finite band splitting changes the condition for a particle-hole excitation to have small energy from $v_Fq\cos(\theta_k-\theta_q)=0$ to $v_Fq\cos(\theta_k-\theta_q)\simeq\pm2|\bm d|$ in inter-band channel. The scattering angle associated with such processes avoids the zeros of the transverse vertex, thereby adding an additional damping term to the underdamped behavior of the transverse nematic mode in the zero-background situation. Qualitatively, compared with the zero-background case, the relative momentum of the electron-hole excitation is shifted by $|\bm{d}|/(v_F q)$; because the decay probability contains the squared vertex, the induced damping term is quadratic in the background $\mathcal{Q}$. Now the propagator of one transverse nematic mode is given by
\begin{equation}
 D_t^{-1}(\Omega,\bm q\,;\mathcal Q)=\underbrace{\rho+q^2+
       \frac{4\bar{\nu}}{v_F^2}\frac{\Omega^2}{q^2}}_{D_{t,\mathcal{Q}=0}^{-1}}
       +\underbrace{\frac{32\bar\nu|\mathcal Q\bm e_{\theta_q}|^2|\Omega|}
                    {v_F^3q^3}}_{\delta D_t^{-1}}.
\label{eq:shift}
\end{equation}
Only one of the transverse modes acquires an additional damping term, originating from interband scattering processes, whereas the smooth intraband velocity associated with intraband scattering merely renormalizes the coefficient of the original term. Consequently, the degeneracy between the two originally degenerate transverse modes is lifted, and this transverse mode with the additional damping term ultimately yields a singular nonanalytic term in the free energy.

For convenience, we reparametrize the order-parameter tensor $\mathcal{Q}$. We define the invariant amplitude $R=(\mathcal{Q}_{\mu i}\mathcal{Q}_{\mu i})^{1/2}$. Let $s_1\ge s_2\ge0$ be the singular values of $\mathcal{Q}$ and define $\eta=(s_1^2-s_2^2)/R^2$, so that $\eta=1$ corresponds to $\alpha$ and $\eta=0$ to $\beta$. That is, $R$ controls the amplitude of the order parameter, whereas $\eta$---which we call the shape parameter---controls its configuration. We further define an important momentum scale $q_R$ and a dimensionless parameter $s$ as follows:
\begin{equation}
 q_R=\left(\frac{16\sqrt{\bar{\nu}}\,R^2}{v_F^2}\right)^{1/3},
 \qquad s=\frac{\rho}{q_R^2}\sim(q_R\xi)^{-2}.
\label{eq:scales}
\end{equation}
where the $q_R$ is the characteristic momentum scale of the fluctuations contributed by the finite background and $\xi\sim 1/\sqrt{\rho}$ is the correlation length indicating how close the system is to the critical point.
The leading singular free-energy density can be written as a scaling function
\begin{equation}
 \delta F_{\mathrm{sing}}(R,\eta;\rho)=
 -\frac{v_Fq_R^4}{8\pi^2\sqrt{\bar{\nu}}}\,
                    \mathfrak F(s,\eta) .
\label{eq:scaling}
\end{equation}
The positive dimensionless function $\mathfrak F$ is derived from the soft-mode determinant in a given nematic background $(R,\eta)$ (see Supplemental Material for details). It satisfies the scaling relation $\delta F_{\mathrm{sing}}[\lambda R,\eta;\lambda^{4/3}\rho]=\lambda^{8/3}\delta F_{\mathrm{sing}}[R,\eta;\rho]$ for $\lambda>0$. Its two asymptotic limits are
\begin{align}
 &q_R^4\,\mathfrak F(s,\eta)\ \sim \,q_R^4\times
 \begin{cases}
 A_{8/3}(\eta)\sim R^{8/3},&s\to0,\\
 s^{-1/4}A_3(\eta)\sim R^3,&s\to\infty,
 \end{cases}\\
 &A_p(\eta)= \int_0^{2\pi}\frac{\mathrm{d}\theta}{2\pi}\big[1+\eta\cos(4\theta)\big]^{p/2}
\end{align}

\begin{figure*}[t]
  \centering
  \includegraphics[width=0.8\textwidth]{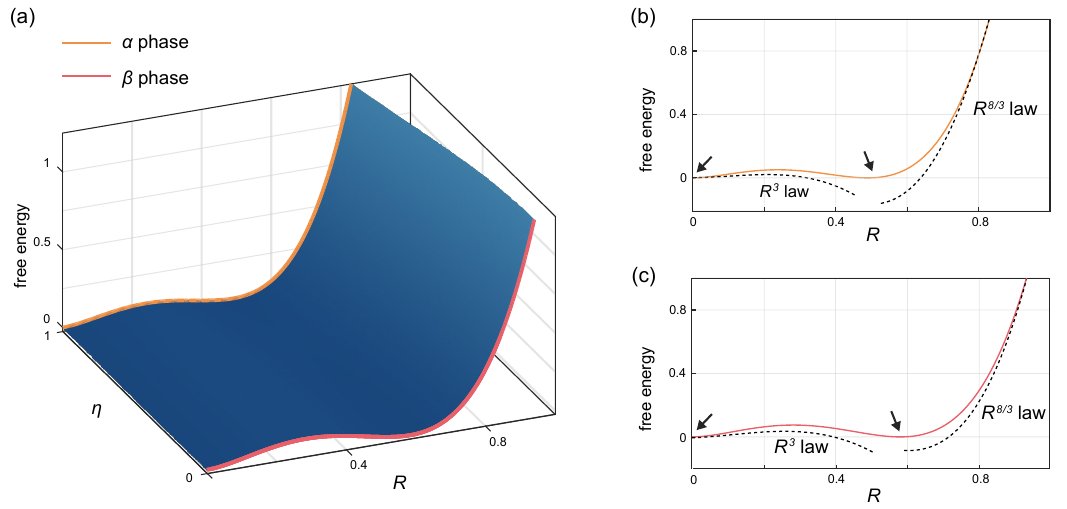}%
  \caption{(a) Free-energy surface in the space of the order-parameter amplitude $R$ and shape parameter $\eta$, corresponding to the three-phase coexistence point. (b) and (c) Free-energy curves of the $\alpha$ and $\beta$ phases, respectively; black arrows mark the presence of local saddle points. The dashed curves show the free energy obtained when the full scaling function is replaced by its asymptotic form in the free-energy correction.}
  \label{Fig4}
\end{figure*}

We stress that the different free-energy corrections considered here---the $R^{8/3}$ and $R^3$ laws---are distinct limits of a general scaling function, rather than two terms that can coexist. In principle, the full scaling function should be employed for the free-energy correction; nevertheless, these two asymptotic limits imply the existence of two qualitatively distinct critical regimes in the phase diagram. Using the full scaling function in Eq.~\eqref{eq:scaling}, we fix $\rho$ and numerically compute the free-energy function $F(R,\eta)$ at the three-phase coexistence point, where the normal, $\beta$, and $\alpha$ phases coexist, as shown in Fig.~\ref{Fig4}. The cuts of $F(R,\eta)$ at $\eta=0$ and $\eta=1$ correspond to the free-energy curves of the $\beta$ and $\alpha$ phases, respectively. We note that, at this point, $R=0$ and a finite value of $R$ are degenerate saddle points.

To elucidate the physical meaning of these two limiting regimes, we examine the relative dynamical correction to the nematic propagator induced by a finite background,
\begin{equation}
r=\frac{\delta D_t^{-1}}{D_{t,\mathcal{Q}=0}^{-1}}\Bigg|_{\Omega=\Omega_t(q)}
\sim \frac{q_R^3}{q^2\sqrt{\rho+q^2}},
\end{equation}
where $\Omega_t=q\sqrt{\rho+q^2}$ is the characteristic frequency of transverse nematic fluctuations in the absence of the background. We compare the two momentum scales $q_R$ and $\sqrt{\rho}$, in order to determine how their relative magnitude is linked to the size of the relative correction $r$. Note that at $q=\sqrt{\rho}$, $r\sim s^{-3/2}$; hence, for $s\gg1$, $\sqrt{\rho}\gg q_R$, and vice versa. We further focus on the characteristic momentum scale $q^*$ at which $r\approx1$, since this scale ultimately dictates the phase-space range---i.e., the upper limit of the frequency-momentum integral---and hence the power law of the free-energy correction. For $s\gg1$, $\sqrt{\rho+q^2}\approx\sqrt{\rho}$ in the denominator of $r$, so setting $r\approx1$ gives $q^*\sim q_R^{3/2}\rho^{-1/4}$. At this momentum, the characteristic frequency of transverse nematic fluctuations is $\Omega_t\sim q_R^{3/2}\rho^{1/4}$, so the phase-space restriction contributes $(q^*)^2\Omega_t\sim q_R^{9/2}\sim R^3$ to the integral. For $s\ll1$, $\sqrt{\rho+q^2}\approx q$ in the denominator of $r$; hence $r\approx1$ yields $q^*\sim q_R$, and correspondingly the phase-space restriction contributes $(q^*)^2\Omega_t\sim q_R^4\sim R^{8/3}$. In summary, a competition arises between the momentum scales $q_R$ and $\sqrt{\rho}$: the former is induced by the finite background, whereas the latter is ``bare'' and drives the system toward the QCP. This competition determines the extent of the low-energy phase space and, in turn, the power law of the free energy.

\section{Structure of the fluctuation-corrected phase diagram}

One may wonder why the positive-definite correction to the nematic propagator, $\delta D_t^{-1}$, yields a negative free-energy correction. This arises from the subtraction of the second-order contribution. The free-energy correction originates from integrating the Gaussian determinant $\ln(1+u)$---where $u$ is the relative dynamical correction $r$ defined above, but with the frequency $\Omega$ treated as the integration variable---and we must subtract $u$, because $u$ is quadratic in the order parameter and merely renormalizes the bare mass $\rho$, thus not contributing to the singular free-energy correction. In other words, the free-energy correction actually comes from $\ln(1+u)-u<0$. Such a decrease in the free energy allows the system to jump directly to a nonzero amplitude before the quadratic coefficient vanishes, thereby turning the second-order transition into a first-order one. 

We also note that $\ln(1+u)-u$ is a concave function, which leads to two important physical consequences.
\begin{itemize}
    \item \textit{The free-energy correction favors the $\alpha$ phase:} Since $u\sim |\bm d_{\theta}|^2$, although both the $\alpha$ and $\beta$ phases satisfy $\int_\theta|\bm d_{\theta}|^2=R^2/2$, the $\alpha$ phase has nodes whereas the $\beta$ phase is a nodeless, uniform gap; that is, for the $\alpha$ phase $u$ is inhomogeneous on the FS, while for the $\beta$ phase it is uniform. The concavity of $\ln(1+u)-u$ makes the inhomogeneous $u$ on the FS yield a more negative integral contribution; thus, the free-energy correction is more negative for the $\alpha$ phase. This significantly enlarges the region of the $\alpha$ phase in the phase diagram compared with the mean-field phase diagram~\cite{sun2008}, as shown in Fig.~\ref{Fig1}.
    \item \textit{The free-energy correction does not induce additional ground-state configurations:} We prove in the Supplemental Material that, because $\ln(1+u)-u$ is concave,
    it is also concave in the shape parameter $\eta^2$. A concave free-energy correction plus the term $u_2$ linear in $\eta^2$ in the original free energy implies that the total free energy cannot possess an additional stable saddle point in the range $0<\eta<1$. This is also corroborated by the numerically computed free-energy surface in Fig.~\ref{Fig4}: no additional saddle point appears along the $\eta$ direction.
\end{itemize}

\section{Experimental outlook and conclusion}

Electronic spectroscopy can first establish which ordered state appears. For initially coincident FS, the $\alpha$ phase has collinear internal polarization and nodal band splitting, whereas the $\beta$ phase has a winding polarization and angle-independent splitting. Spin-resolved photoemission can distinguish these patterns in a real-spin realization, including the internal order of $\beta$ that circular Fermi contours alone cannot reveal. Measurements on metallic $\mathrm{KV_2Se_2O}$ have resolved a $d$-wave spin texture~\cite{jiang2025}, while photoemission studies of $\mathrm{MnTe}$ and $\mathrm{CrSb}$ have mapped momentum-dependent band splitting~\cite{krempasky2024,yang2025,ding2024}. These materials provide spectroscopic benchmarks; establishing the present mechanism also requires a tunable quadrupolar instability. Tracking the splitting and polarization across such a transition would identify the selected state and reveal whether its onset is discontinuous.

The static susceptibility at phase coexistence provides a quantitative experimental discriminator between the two critical regimes. Let $\rho_a^*$ denote the coexistence point between the normal state and the ordered state $a=\alpha,\beta$. Within either asymptotic power-law regime, the static nematic susceptibilities satisfy
\begin{equation}
 \frac{\chi_a(\rho_a^*)}{\chi_N(\rho_a^*)}
 = \frac{1}{p-2},
 \qquad a=\alpha,\beta ,
 \label{eq:response}
\end{equation}
where $p=8/3$ or $3$ is the power of the nonanalytic free-energy correction. Here, $\chi_a$ and $\chi_N$ are the amplitude responses within the ordered and normal phases, evaluated with the same normalization of the conjugate field. The ratio is $3/2$ for $p=8/3$ and $1$ for $p=3$. This $50\%$ difference in the normalized response distinguishes the two asymptotic regimes; outside these limits, the response must be obtained from the full scaling function.

\begin{figure*}[t]
  \centering
  \includegraphics[width=0.9\textwidth]{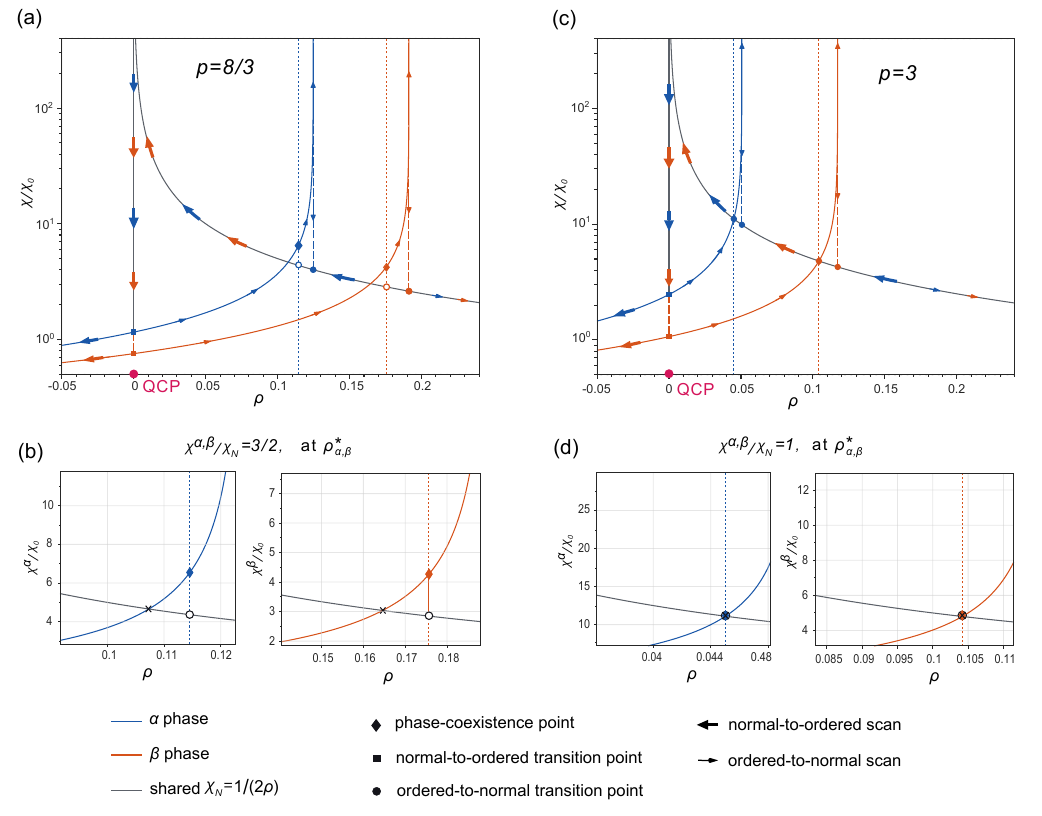}
  \caption{(a) Hysteresis loops of the nematic susceptibility in the $p=8/3$ critical regime, with the vertical axis on a logarithmic scale. Sweeping $\rho$ from positive to negative values corresponds to a normal-to-ordered scan, while the reverse sweep corresponds to an ordered-to-normal scan. The first-order transition produces hysteresis in the susceptibility, with distinct divergence points (transition points) for the two sweep directions. The solid gray line denotes the common segment of the $\alpha$- and $\beta$-phase sweep paths and defines the reference curve $\chi_N$, the normal-phase susceptibility. The dashed red and blue lines mark the coexistence points $\rho_a^*$ between the corresponding ordered phase $a$ and the normal phase. (b) Enlarged views near the two-phase coexistence points for the $\alpha$ phase (left) and the $\beta$ phase (right). At coexistence, the susceptibility has a fixed ratio to the reference value $\chi_N$. (c), (d) Corresponding results in the $p=3$ critical regime.}
  \label{Fig5}
\end{figure*}

Measurements of nematic-susceptibility hysteresis provide a practical route to this criterion, as illustrated in Fig.~\ref{Fig5}. The first step is to locate $\rho_a^*$ through thermodynamic measurements, including latent-heat measurements along a finite-temperature first-order boundary where available. The coincident normal-state portions of the $\alpha$- and $\beta$-phase scans then determine a common reference curve $\chi_N(\rho)$, whose value at $\rho_a^*$ supplies the baseline. Comparing the response within the ordered phase at coexistence with this reference yields the ratio in Eq.~\eqref{eq:response}. Measurements approaching zero temperature would allow comparison with the asymptotic predictions.

Several experimental approaches already support this strategy. Elastoresistivity has tracked nematic susceptibility across iron-based superconductors~\cite{chu2012,kuo2016}, including $\mathrm{FeSe}_{1-x}\mathrm{S}_x$~\cite{hosoi2016}, while strain tuning has provided continuous control of proximity to nematic quantum criticality~\cite{worasaran2021}. The coupling between nematic order and lattice strain renormalizes the shear modulus, making shear-modulus measurements an indirect probe of the nematic susceptibility~\cite{bohmer2014}; elastocaloric measurements complement this probe through the associated entropy response~\cite{ikeda2021}. Applying these methods to the present order requires identifying a field that couples linearly to the relevant component of $\mathcal Q$ and calibrating the measured response in both phases. For the (pseudo-)spin-channel ordering, strain alone generally does not supply this conjugate field.

The nematic spectrum offers a complementary test of the feedback underlying QOBD. In the normal state, the transverse mode is underdamped, with $\omega_t\propto q^2$ in the critical momentum range, whereas the longitudinal response is overdamped. A nematic background opens an additional decay channel for one transverse branch, as described by Eq.~\eqref{eq:shift}. At fixed momentum, while the peak remains weakly damped, its added linewidth follows the squared electronic band splitting. The leading angular contribution therefore vanishes along the nodes of $\alpha$ and is angle independent in $\beta$. Comparing the linewidth before and after ordering, while varying the momentum direction, would test the decay mechanism responsible for the singular free-energy gain.

Existing measurements provide a starting point for such spectroscopy. Polarization-resolved Raman scattering has detected charge-nematic fluctuations in iron arsenides and $\mathrm{FeSe}$~\cite{gallais2013,massat2016}, and followed the evolution of quadrupolar fluctuations into a coherent mode below the superconducting transition in $\mathrm{NaFe}_{1-x}\mathrm{Co}_x\mathrm{As}$~\cite{thorsmolle2016}. Combining Raman scattering with controlled strain has further revealed how nematic spectra change under an applied field~\cite{philippe2022}. These experiments establish access to nematic dynamics in selected symmetry channels. Their nearly zero momentum transfer, however, does not resolve the finite-momentum dispersion required here, and the measured charge response differs from the spin- or pseudospin-dependent response of the present model.

At finite momentum, spin-polarized electron energy-loss spectroscopy has measured spin-wave dispersion~\cite{vollmer2003,etzkorn2005,tang2007,prokop2009} and magnon lifetimes~\cite{qin2015,zakeri2012,qin2013magnons} in ultrathin metallic films. Resonant inelastic X-ray scattering has resolved the dispersion and damping of magnetic excitations in detwinned $\mathrm{FeSe}$~\cite{lu2022,liu2024rixs}, while angular analysis has separated spin, charge, and orbital contributions to the scattering signal~\cite{kang2019,zhang2022rixs,morettisala2011}. These capabilities suggest possible routes to the predicted nematic spectrum, provided the probe couples to the relevant quadrupolar mode and has sufficient resolution. In a suitable material, combining these measurements with electronic band splitting would connect the change in nematic damping to the emergence of order.

We have developed a theory linking electronic incoherence and phase selection in a two-component, two-dimensional nematic metal. Its uniform ordered states include a nodal $\alpha$ phase and a $\beta$ phase with winding internal polarization. Near the putative QCP, overdamped longitudinal fluctuations generate the leading $|\omega|^{2/3}$ electronic self-energy, while underdamped transverse fluctuations produce a logarithmic correction. The decisive feedback arises when nematic order changes electronic scattering and opens an additional transverse decay channel. This change generates a negative singular free-energy correction, described by a general scaling function connecting the $R^{8/3}$ and $R^3$ limits. Their different powers reflect the competition between fluctuation momenta set by the normal state and the nematic background. The correction favors $\alpha$, enlarges its stability region, and introduces no stable uniform configuration between $\alpha$ and $\beta$. Its competition with the mean-field energy determines the discontinuous $\alpha$--$\beta$ boundary. The onset of nematic order becomes first order, preempting the mean-field QCP. The logarithmic electronic contribution must therefore be assessed near the actual phase boundary, where its importance cannot be inferred solely from its subleading behavior at criticality. The coexistence susceptibility ratios and the angular pattern of added nematic damping turn this feedback into complementary experimental tests. More broadly, our results show that the significance of a fluctuation channel depends jointly on how it scatters electrons and how it reorganizes the phases those electrons can occupy.

\textit{Acknowledgments}.--- 
This work was financially supported by the National Key R\&D Program of China (Grants
No. 2022YFA1403200 and No.2024YFA1613200), National Natural Science Foundation of
China (Grants No. 92265104, No. 12022413), the Basic Research Program of the Chinese
Academy of Sciences Based on Major Scientific Infrastructures (Grant No. JZHKYPT-2021-08), the CASHIPS Directors Fund (Grant No. BJPY2023A09), Anhui Provincial Major S\&T
Project(s202305a12020005), and the High Magnetic Field Laboratory of Anhui Province under
Contract No. AHHM-FX-2020-02

\bibliographystyle{apsrev4-2}
\bibliography{ref.bib}

\end{document}